\documentclass[sigconf,nonacm]{acmart}
\AtBeginDocument{%
  }

\setcopyright{acmlicensed}
\copyrightyear{2018}
\acmYear{2018}
\acmDOI{XXXXXXX.XXXXXXX}
\acmConference[Conference acronym 'XX]{Make sure to enter the correct
  conference title from your rights confirmation email}{June 03--05,
  2018}{Woodstock, NY}
\acmISBN{978-1-4503-XXXX-X/2018/06}

\usepackage{tabularx}

\begin{document}

\title{Who Is Doing the Thinking? Characterising Human Agency in First-Year Students' AI Interactions}

\author{Ali Shakiba}
\email{ali.shakiba@sydney.edu.au}
\orcid{}
\correspondingauthor
\affiliation{%
  \institution{The University of Sydney}
  \city{Sydney}
  \state{NSW}
  \country{Australia}
}

\author{Thomas Chaffey}
\email{thomas.chaffey@sydney.edu.au}
\orcid{}
\affiliation{%
	\institution{The University of Sydney}
	\city{Sydney}
	\state{NSW}
	\country{Australia}
}
\renewcommand{\shortauthors}{Shakiba et al.}

\begin{abstract}
  When students use generative AI to produce their work, the learning process becomes invisible. Established frameworks such as Bloom's taxonomy then classify the cognitive level of what the AI produced, not what the student contributed. We propose the Human Agency Depth Taxonomy (HADT), a preliminary six-tier taxonomy that classifies the engineering contribution a student makes in a prompt, independent of the AI's output. HADT combines three dimensions: contextual bandwidth, engineering abstraction and student agency. We applied HADT to 962 prompts from 105 students' structured AI Journals in an AI-permissive, first-year software engineering course of 145 students, and examined how agency changed over ten journalled weeks. Most prompts (84.2\%) carried some student contribution beyond delegation. Agency tracked the development lifecycle. Architecture-discovery prompts dominated the requirements phase (48\% of turns) and fell to 5-12\% from week 9, while prompts analysing existing systems rose during build and verification (odds ratio 2.88, with turns clustered by student). Two-thirds of the variance in tier lay within students, so agency depended more on the phase of work than on the student. A turn-level rise in delegation in deadline weeks did not hold once turns were clustered by student. Explicitly specifying constraints on the AI's output was rare throughout, which we interpret as a knowledge bottleneck for novices rather than a gap in instruction. We discuss implications for course and assessment design, and for using HADT as a formative lens on how students engage with AI.
\end{abstract}




\maketitle

	\section{Introduction}\label{sec:intro}
		Generative AI (GenAI) has quickly become a mainstream tool in computing education. Large language models (LLM) now solve typical programming exercises with high accuracy \cite{finnie22robots_are_coming,finnie23_my_ai_wants_to_know_exam,denny24_cse_era_gen_ai}, and students use them across the software development lifecycle (SDLC) \cite{daun23_chatgpt_change_soft_eng,prather23_robots_are_here}, from eliciting requirements to generating designs, code and documentation. In response, many courses have moved from prohibiting AI to permitting, and in some cases teaching its use \cite{shakiba26_making,lau23_from_ban_it_to_resist,lodge23_assessment_reform_age_ai}.
		
		This shift exposes an old assessment problem in a new form. Consider two identical arithmetic results, one worked by hand and the other produced with a calculator. Without seeing the working, an assessor cannot tell them apart. AI-permissive courses face the same difficulty at scale: when only the final artefact is assessed, whether a report, a design or a working application, the process behind it and the student's contribution to it become invisible\cite{shakiba26_making,bearman20_prep_uni_assess_world_with_ai}. Unrestricted AI use carries an additional risk. Students may offload the cognitive work of a task to the model, producing polished artefacts while conceptual gaps go undetected \cite{prather24_widening_gap,bastani25_gen_ai_can_harm_learning,kazemi23_effect_ai_code_gen_novice_learners}. Drawing on Vygotsky's zone of proximal development \cite{vygotsky78_zone_of_proximal}, a zone of no development is created \cite{santos25zone_no_development}: the AI's support is so complete that the task no longer requires the student to engage in the learning it was intended to.
		
		Reflection instruments can make the process visible again. Structured AI journals \cite{shakiba26_making}, in which students log their prompts and outputs and reflect on how they verified them, give educators a record of human-AI interaction. Analyses of such records characterised how novices oversee AI output \cite{shakiba2026beyond_correctness}. Visibility, however, is not a measurement. To determine from these records whether and when students are engaged in the task or delegating it to AI, educators need a way to classify the extent of engineering thinking that students contribute to each interaction.
		
		Existing frameworks are poorly suited to this purpose. Taxonomies such as revised Bloom's \cite{krathwohl02_revised_bloom,thompson08_bloom_cs_taxonomy} and SOLO \cite{biggs14solo,lister06_not_seeing_forest_for_trees} classify the cognitive level of a task or a response. When AI produces the response, these classifications describe the machine's output rather than the student's thinking. Consider a first-year student who prompts: ``Help me design the data structure for our Water Rescue Coordination System in Power Apps. The app needs to store incidents, volunteers, assignments, and rescue equipment ...''. Under Bloom's taxonomy, this is a Create-level task. Yet, the label does not tell us whether the student engaged in any of the designing. They may have specified the entities, constraints and acceptance criteria that shape the answer, or they may have delegated the entire problem to the model. The same label applies in both cases.

        This paper addresses that gap. We propose the \textsc{Human Agency Depth Taxonomy} (HADT), a six-tier taxonomy that classifies the \emph{engineering contribution} a student makes through a prompt, \emph{independent} of what the \emph{AI generates}. HADT combines three dimensions: \textsc{Contextual Bandwidth} (generic to project-specific), \textsc{Engineering Abstraction} (local to system-wide), and \textsc{Student Agency} (whether the student or the AI defines the problem). We apply HADT to 962 prompts from collected AI journals in a 13-week first-year software engineering course of 145 students, in which AI use was permitted and taught. We address two research questions:
		\begin{description}
			\item[RQ1:] What dimensions and levels distinguish the human contribution in students' prompts, independent of what the AI generates?
			\item[RQ2:] How is human agency distributed across first-year students' AI interactions, and how does it change over the semester?
		\end{description}
		This paper makes three contributions:
		\begin{enumerate}
			\item HADT, a taxonomy of human agency in AI prompts, with three dimensions, six tiers, and decision rules.
			\item An empirical account of agency across a semester. Most interactions are not passive (84.2\% above T0). Agency follows the SDLC, from design-space exploration early to implementation and diagnosis later.
			\item Implications for course and assessment design. If agency depends on the situation rather than the student, the timing and structure of assessment shape how students use AI.
		\end{enumerate}
		The rest of the paper is organised as follows: Section \ref{sec:lit} reviews the background and literature. Section \ref{sec:method} describes the method: the course and data, and the development of HADT (RQ1). Section \ref{sec:results} reports agency patterns (RQ2). Section \ref{sec:discuss} discusses implications, Section \ref{sec:limits} sets out limitations and threats to validity, and Section \ref{sec:conc} concludes with future research directions.
		
	\section{Background and Related Work}\label{sec:lit}
        The widespread availability of GenAI has challenged established approaches to teaching and assessing programming and software engineering. Early studies demonstrated that code-generating models could solve most typical introductory programming exercises and examination questions \cite{finnie22robots_are_coming}. Subsequent research has identified opportunities for personalised assistance, feedback, and exercise generation, alongside concerns about academic integrity, over-reliance, and the validity of traditional assessments \cite{denny24_cse_era_gen_ai,prather23_robots_are_here}. In response, educators have adopted approaches ranging from restricting AI use to redesigning learning activities and assessments around its capabilities \cite{lau23_from_ban_it_to_resist,shakiba26_making}.

        A central concern is whether improved performance on AI-assisted tasks translates into meaningful learning. Empirical findings remain mixed. In a controlled study, novice programmers using an AI code generator completed more tasks without a corresponding decline in subsequent unaided performance \cite{kazemi23_effect_ai_code_gen_novice_learners}. However, observational research has documented students accepting generated solutions without fully understanding them, highlighting cognitive and metacognitive difficulties in AI-assisted programming \cite{prather23_it_knows_what_i_want_copilot}. Prather et al. further observed that GenAI could widen differences between learners: students with stronger metacognitive skills benefited from the assistance, whereas others developed an illusion of competence \cite{prather24_widening_gap}. Similar concerns extend beyond computing education. A large field experiment found that unrestricted chatbot access improved practice performance but reduced subsequent unassisted performance, whereas pedagogically constrained assistance largely mitigated this effect \cite{bastani25_gen_ai_can_harm_learning}. Together, these findings suggest that successful task completion is not necessarily evidence of learning and that how students engage with AI assistance matters.

        This distinction can be understood through the literature on cognitive offloading and task delegation. Cognitive offload refers to the use of external resources to reduce cognitive demands, which can free up capacity for other activities while also reducing involvement in important learning processes \cite{RiskoGilbert16}. Research on task delegation has examined the activities people prefer to assign to AI systems and the factors that influence these decisions \cite{LubarsTan19}. More recent studies have associated GenAI use with reduced critical-thinking effort among knowledge workers \cite{Lee25} and with patterns of metacognitive disengagement in educational settings, where improvements in task performance are not necessarily accompanied by comparable gains in knowledge or transfer \cite{Fan25,Gerlich25,prather24_widening_gap}. Although these findings do not establish that delegation is inherently detrimental, they highlight the importance of distinguishing between AI assistance that supports students' reasoning and assistance that substitutes for it.

        One way to examine this distinction is through students' interactions with GenAI systems, particularly their prompts. Existing prompting research has focused primarily on techniques for eliciting effective model responses. Prompt-pattern catalogues and surveys describe strategies such as few-shot examples, role prompting, and chain-of-thought prompting \cite{White23,Schulhoff24}. In computing education, Prompt Problems have been proposed to develop and assess students' ability to formulate prompts that produce correct code \cite{Denny2024Prompt}. Other studies have analysed student-AI conversations by classifying prompts according to their conversational function, such as initial requests, follow-up questions, and corrections, or according to the problem-solving issues they address \cite{Scholl24}. Prompts have also been mapped to cognitive levels using the revised Bloom's taxonomy \cite{Chen25}. These approaches provide useful insights into prompting strategies, interaction patterns, and the cognitive demands of requested tasks. However, they do not necessarily distinguish between prompts that delegate an entire problem to AI and those that communicate substantial reasoning or problem-solving work already undertaken by the student. A technically sophisticated prompt, for example, may specify detailed output requirements without providing evidence that the student has independently analysed the underlying engineering problem.

        Established educational taxonomies offer complementary perspectives on cognitive engagement and learning. The revised Bloom's taxonomy classifies cognitive processes from remembering to creating \cite{krathwohl02_revised_bloom}. It has been widely applied to programming assessment design, although difficulties in consistently interpreting its categories within computing have motivated domain-specific adaptations \cite{johnson06_is_bloom_appropriate_cs,thompson08_bloom_cs_taxonomy}. The SOLO (Structure of Observed Learning Outcomes) taxonomy instead examines the structural complexity of learners' responses \cite{biggs14solo}, including their ability to read, interpret, and explain programs \cite{lister06_not_seeing_forest_for_trees,whalley06_australasian_study_reading_comprehension_novice_programmers}. The ICAP (Interactive, Constructive, Active, Passive) framework approaches learning through observable engagement, distinguishing passive, active, constructive, and interactive behaviours and proposing that deeper forms of engagement generally support stronger learning outcomes \cite{chi14_icap_framework}.

        While these frameworks remain valuable, applying them to AI-assisted work raises a fundamental question: whose cognitive contribution is being observed? Bloom's taxonomy describes the cognitive demands of a task, but a prompt asking AI to design a database schema does not establish that the student performed the associated design reasoning. Similarly, the structural complexity of an AI-generated response does not necessarily reflect the student's understanding. ICAP offers a closer perspective by focusing on learner engagement, but its categories were not developed specifically to distinguish the intellectual contributions of a learner and a generative system. For example, submitting an assignment brief to a chatbot is an observable action, yet provides little evidence that the student has independently framed or analysed the underlying engineering problem. Consequently, neither the cognitive level of the requested task, the sophistication of the generated response, nor the observable interaction alone necessarily reveals the extent of the student's contribution.

        Recent research has begun adapting established cognitive frameworks to AI-assisted education, including proposals for AI-specific competencies \cite{gonsalves26_gen_ai_ct} and approaches that use Bloom's taxonomy to guide the integration of GenAI into software engineering and computing curricula \cite{garousi2026implementinggenaisupportedlearningsoftware}. Such work provides guidance on the competencies students should develop and the types of activities in which AI assistance may be appropriate. However, a complementary need remains for methods that characterise how students distribute problem-solving responsibilities between themselves and AI during actual interactions.

        Our proposed HADT characterises the student's observable engineering contribution within a prompt, rather than the sophistication of the prompting technique, the cognitive level of the requested task, or the quality of the generated output. This distinction is important: \emph{a carefully structured prompt containing few-shot examples and explicit formatting instructions may still delegate the underlying engineering problem almost entirely to AI, whereas a brief prompt presenting the student's own analysis, evidence, or hypothesis may demonstrate substantial prior problem framing}. By examining what students contribute before requesting assistance, HADT provides a prompt-level perspective on the distribution of engineering work between students and GenAI. It does not directly measure learning, cognitive effort, or the correctness of students' reasoning; rather, it provides a means of characterising observable patterns of agency and delegation that can inform the interpretation of AI-assisted student work.
	
	\section{Method}\label{sec:method}
		\subsection{Context and Data}\label{sec:context}
            The study took place in ELEC1005 Introduction to Software Engineering at the University of Sydney, a first-year, first-semester course. The cohort comprised 145 students, 140 of whom were in their first semester at university, and no prior programming experience was assumed. The course provides a compressed introduction to the entire SDLC, from planning and requirements through design and implementation to verification and deployment, structured around a team project spanning two assignments (A1 and A2). Rather than teaching programming, the course focuses on engineering processes and uses industry tools: Jira for planning, Figma for prototyping, Git for version control, UML diagrams for design, and a low-code/no-code environment (Power Apps with SharePoint lists) for implementation.
            \begin{table*}[htbp] 
                \caption{Weekly lecture topics, project activities, assessments, AI instruction, and analysis phases.} 
				\label{tab:weekly-schedule} 
				\centering \small \renewcommand{\arraystretch}{1.2} 
				\begin{tabularx}{\textwidth}{ >{\centering\arraybackslash}p{0.04\textwidth} >{\raggedright\arraybackslash}X >{\raggedright\arraybackslash}p{0.16\textwidth} >{\raggedright\arraybackslash}p{0.17\textwidth} >{\raggedright\arraybackslash}p{0.22\textwidth} >{\raggedright\arraybackslash}p{0.15\textwidth}}
					\toprule \textbf{Wk} & \textbf{Lecture topic} & \textbf{Group Project} & \textbf{Assessment} & \textbf{Explicit AI instruction} & \textbf{Analysis phase} \\ 
					\midrule 
					1 & Intro to SDLC & --- & --- & --- & Not journalled \\
					2 & Planning \& Agile (Jira) & Groups form & --- & GenAI and agency & Not journalled \\ 
					3 & Analysis \& UX, user stories, Figma & A1 starts & Early-feedback quiz (AI allowed) & AI Journal intro + warm-up; GenAI for Requirements engineering (``context starvation'') & Excluded (warm-up) \\ 
					4 & Requirements engineering, Functional and Non-functional requirements, acceptance criteria, Git & A1 requirements & --- & LLMs and Requirements engineering & Analysis \\ 
					5 & Architecture \& design, UML & A1 design & --- & --- & Design \& prototype \\ 
					6 & Implementation: manual $\rightarrow$ no-code $\rightarrow$ AI coding & A1 prototype & Mid-sem quiz & AI-assisted coding & Design \& prototype $\cdot$ deadline \\ \midrule \multicolumn{6}{c}{\textit{Mid-semester break}} \\ \midrule 
					7 & Design patterns, Object-oriented design & A1 due Fri; A2 released & A1 due & --- & Design \& prototype $\cdot$ deadline \\ 
					8 & Verification \& validation & A1 demos; A2 build & A1 demos & --- & Build \& verify \\ 
					9 & Quality engineering, CI/CD & A2 build & --- & --- & Build \& verify \\ 
					10 & CI/CD, documentation & A2 build; V\&V practical & --- & --- & Build \& verify \\ 
					11 & System design \& architecture, Architectural decision records & A2 build & --- & --- & Build \& verify \\ 
					12 & Technical debt, spec-driven development & A2 deliverables & A2 due & Spec-driven AI across the SDLC & Build \& verify $\cdot$ deadline \\ 
					13 & Career paths, wrap-up & A2 demos; final deliverables Fri & A2 demos & --- & Build \& verify $\cdot$ deadline \\ 
					\bottomrule 
				\end{tabularx} 
			\end{table*}
            Students declared their AI use through a structured AI Journal \cite{shakiba26_making}, which records each human-AI conversation in two layers. The first layer, execution tracking, captures the prompt, the AI's output, the model and version, the environment, and the student's intent. The second layer, cognitive auditing, asks students to reflect on how they verified the output and their confidence in that verification (1-5), how useful the AI was (1-5), what they changed or rejected, their most important learning moment, and which parts would be hardest to complete again without AI. This paper analyses only the prompts captured in the first layer.
            
			AI and human agency were introduced in Week 2 by sharing real-world use cases of inresponsible use of AI as comic books and an in-class padlet activity. In Week 3, students were introduced to the AI Journal, worked through real-world case studies, and completed a warm-up journal, which was excluded from the analysis. From Week 4 onwards, each student submitted a weekly declaration, either reporting that they had not used AI that week or providing an AI Journal documenting their use. The journal was an unmarked hurdle task. Students were explicitly told that their journals would never be used to reduce their marks and could instead serve as evidence of their own contribution, for example, in team disputes. This design was intended to encourage honest disclosure rather than responses optimised for assessment.
			
\begin{figure}
				\centering
				\includegraphics[width=0.9\columnwidth]{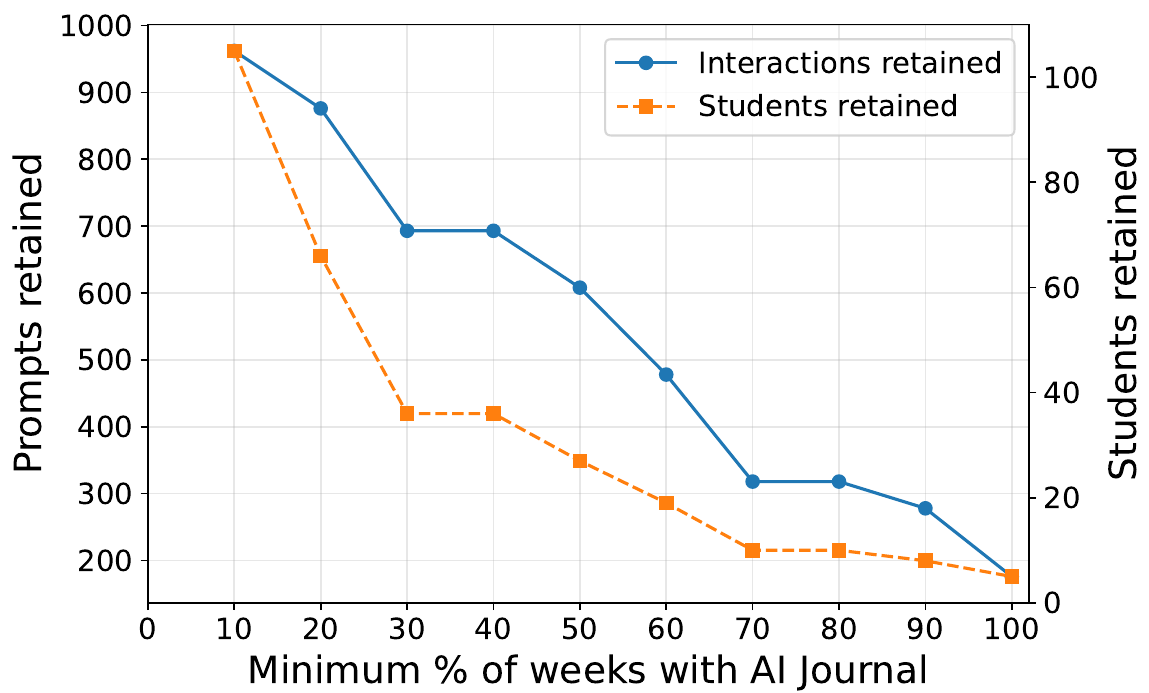}
				\caption{Prompts and students retained by minimum share of weeks with an AI Journal; the 50\% threshold defines the trajectory subset (27 students, 608 turns) used in the analysis of Section \ref{sec:within-student}}
				\label{fig:retained}
			\end{figure}

			The dataset comprises all 962 prompts submitted through AI Journals between Weeks 4 and 13 by 105 students (40 students declared they did not use AI at all). These prompts were coded and are analysed in Section \ref{sec:results}. The number of submitted prompts declined over the semester, from 187 in Week 4 to between 58 and 75 per week from Week 8 onwards (Table \ref{tab:overall-dist}). For the longitudinal analysis (Section \ref{sec:within-student}), we retain the 27 students who submitted an AI Journal in at least 50\% of the eligible weeks, contributing 608 prompts in total (Figure \ref{fig:retained}).
			
		\subsection{Human Agency Depth Taxonomy}\label{sec:hadt}
			HADT classifies each prompt according to the engineering contribution the student makes through it, independent of what the AI generates. We initially coded prompts using the revised Bloom's taxonomy \cite{krathwohl02_revised_bloom} as part of the course's routine weekly quality assurance. This approach quickly encountered the measurement problem described in Section 1. Prompts such as the database-schema example could be classified as create, yet they ranged from fully delegated requests to tightly constrained specifications in which the student had already defined much of the problem. We therefore set Bloom's taxonomy aside and examined what distinguished prompts that contained evidence of student thinking from those that did not.
			
			Three recurring differences emerged: the amount of project-specific context supplied by the student, the level of the system addressed by the request, and whether the student or the AI defined the problem. We treated these as the three dimensions of the proposed taxonomy. The first author then grouped recurring combinations of these dimensions into tiers. The tier boundaries and decision rules were refined over two rounds of coding until additional prompts could be classified without requiring further revisions to the scheme.
			
			Each dimension describes one aspect of what the student, rather than the AI, contributes:
			\begin{description}
				\item[Contextual bandwidth] \textbf{(generic to project-specific).} The amount of project-specific context the student contributes, including requirements, data, constraints, earlier decisions, or evidence. A generic prompt (``What is a primary key?'') could come from any student. A project-specific prompt contains information specific to the system being built. High contextual bandwidth provides evidence that the student is bringing knowledge of their project into the interaction and making that knowledge available to the AI.
				\item[Engineering abstraction] \textbf{(local to system-wide).} The scope at which the request engages with the system, ranging from an isolated component (a formula, a screen, or a table) to that component in the context of the wider system (its interfaces, data flows, requirements, and trade-offs). System-wide prompts provide evidence that the student is considering how multiple parts of the system fit together, a core software engineering competence.
				\item[Student agency] \textbf{(AI defines the problem to student defines the problem).} The extent to which the student, rather than the AI, frames the problem. At the low end, the student forwards a task, often by pasting an assignment brief, lab sheet, or marking criteria, and leaves the AI to interpret, decompose, and solve it. At the high end, the student has already framed and decomposed the problem, specifying what needs to be addressed, and uses the AI to execute, diagnose, or explore within that frame.
			\end{description}
            The dimensions are related but not redundant, and no single dimension increases monotonically across the tiers. A prompt can be highly project-specific yet exhibit low student agency, for example when a student pastes the full project brief and asks the AI to solve it. Similarly, a T4 diagnostic request may concern a single local formula and therefore have lower engineering abstraction than a T3 request framed by system-wide constraints. HADT therefore defines each tier by the combined profile of the three dimensions and the engineering act requested: retrieving knowledge, generating a component, generating under constraints, diagnosing an existing artefact, or exploring the design space (Table \ref{tab:hadt-profile}). The tiers are ordered by the extent of engineering reasoning evident in the student's prompt before the AI responds. A diagnostic prompt ranks above constrained generation because the student contributes an existing artefact or observation together with a hypothesis about its cause. Design-space exploration ranks highest because the student frames the alternatives or constraints to be explored while retaining responsibility for the design decision.
			\begin{description}
				\item[T0 - Delegating the problem.] The prompt forwards a task without any student-authored constraints, decomposition, or evidence. \\ \textbf{Rule:} If pasted material (e.g., an assignment brief, lab sheet, or marking criteria) were removed, no student contribution to framing or solving the problem would remain.
				\item[T1 - Looking up knowledge.] The prompt asks a factual or how-to question (e.g., a definition, syntax, or tool feature) that is independent of the student's project. \\ \textbf{Rule:} The answer would be essentially the same for any student.
				\item[T2 - Requesting components.] The prompt requests a specific, local artefact (e.g., a formula, function, screen, or table) and states the student's intent for that artefact, but not the wider system constraints it must satisfy. \\ \textbf{Rule:} The student has decomposed the problem into a specific component, and the request is scoped to that component.
				\item[T3 - Specifying constraints.] The prompt requests an artefact and specifies the system constraints it must satisfy (e.g., requirements, interfaces, data model, or acceptance criteria). \\ \textbf{Rule:} The constraints come from the student's own system and provide criteria against which the output can be evaluated.
				\item[T4 - Analysing existing systems.] The student supplies evidence from a system they have built (e.g., an error, output, or artefact), together with a hypothesis or question, and asks the AI to diagnose or evaluate it. \\ \textbf{Rule:} The student brings an existing artefact or observation to the interaction and asks the AI for diagnosis or evaluation rather than replacement.
				\item[T5 - Constraint-driven architecture discovery.] The student frames alternative designs, scenarios, or ``what-if'' constraints and uses the AI to explore the design space and its trade-offs. \\ \textbf{Rule:} The student frames the alternatives or constraints to be explored, and the design decision remains with the student.
			\end{description}
                \begin{table*}[t]
                \caption{Dimension profile for each HADT tier. The dimensions are not individually monotonic across tiers; tiers are distinguished by their dimension profile together with the engineering act requested.}
                \label{tab:hadt-profile}
                \centering\small
                \begin{tabularx}{\textwidth}{@{}p{0.05\textwidth}p{0.27\textwidth}p{0.20\textwidth}p{0.20\textwidth}X@{}}
                    \toprule
                    \textbf{Tier} & \textbf{Contextual bandwidth} & \textbf{Engineering abstraction} & \textbf{Student agency} & \textbf{Engineering act}  \\
                    \midrule
                    T5 & High (own requirements, constraints) & System-wide & Frames alternatives; keeps decision & Explore \\
                    T4 & High (evidence from own artefact) & Local or system-wide & Frames a hypothesis & Diagnose / evaluate  \\
                    T3 & High (own requirements, data model) & System-wide & Fixes the constraints & Generate under constraints  \\
                    T2 & Medium (names own component) & Local & Has decomposed the problem & Generate component \\
                    T1 & Low (generic) & Local / none & Narrow factual question & Retrieve  \\
                    T0 & Low (teaching material only) & As given by brief & AI defines the problem & Delegate  \\
                    \bottomrule
                \end{tabularx}
            \end{table*}
		
		\subsection{Coding Procedure}
			The unit of analysis is a single prompt, defined as one student turn within an AI Journal entry. An AI Journal entry records a conversation containing one or more student prompts, each of which was coded separately and assigned exactly one HADT tier.
		
            Coders used only the prompt text; other journal fields were not consulted. Content within the prompt was attributed to the student only when it went beyond material provided in the course, such as the student's own requirements, designs, data, decisions, or evidence from their system. A prompt that reproduced or closely followed teaching material, such as an assignment brief, lab sheet, or marking criteria, without additional student-provided context was coded as T0, regardless of its length. This rule prevents copied task context from being treated as evidence of student engineering contribution.
            
            We used the following decision rules for edge cases:
            \begin{itemize}
				\item Pasted teaching material with no student-authored addition is assigned T0.
				\item Multi-part prompts are assigned the highest tier supported by their student-authored content.
				\item Follow-up prompts are coded independently using only the text of the current prompt, without reference to earlier prompts in the conversation. A follow-up that itself supplies evidence (e.g., ``this formula returns blank when the volunteer list is empty'') can reach T4; a bare ``fix it'' or ``try again'' adds no student engineering contribution and is assigned T0. This rule reflects what HADT measures: the \emph{engineering contribution} added by each prompt. Context already available to the AI from earlier prompts or conversation history is not treated as a new contribution in the current prompt. A short follow-up that states a decision (e.g., ``key it on volunteers, not incidents'') is still credited because the decision is explicit in its text.
				\item A reference to an attachment contributes to contextual bandwidth only when the prompt explicitly refers to the attachment and indicates what it contains or why it matters (e.g., ``using the attached ER diagram, check whether the Assignments table supports multiple volunteers''). A bare ``see attached'' does not contribute to contextual bandwidth, and an attachment alone never raises the tier.
			\end{itemize}
			The first author coded all prompts blind to student identity. To assess inter-rater reliability, the second author independently coded a random sample of approximately 27\% of the prompts ($n = 261$), using Table \ref{tab:hadt-profile} and the decision rules above. Because HADT tiers are ordinal, we report linearly weighted Cohen's $\kappa$ \cite{cohen1968weighted} as the primary reliability measure, alongside unweighted Cohen's $\kappa$ \cite{cohen1960coefficient} and percentage agreement. We interpret $\kappa$ values using the Landis and Koch bands \cite{landis1977measurement}.

            The coders agreed on 87.4\% of the 261 double-coded prompts. Unweighted Cohen's $\kappa$ was 0.82 (95\% CI $0.77-0.88$), and linearly weighted Cohen's $\kappa$ was 0.88 (95\% CI $0.84-0.92$), indicating almost perfect agreement under the Landis and Koch interpretation. Agreement was complete for T0, T1, and T4. Of the 33 disagreements, 30 occurred between T3 and T5; in 29 of these, the first author assigned T5 and the second author T3. These prompts asked the AI to design an artefact from student-supplied requirements or constraints, creating ambiguity between constrained generation (T3) and design-space exploration (T5). The remaining three disagreements were between T2 and T3.

            Both coders were also course co-coordinators and lecturers: the first author taught primarily in the second half of the course and the second author in the first half. Their familiarity with the course context may have informed their interpretation of the prompts, although both applied the same documented taxonomy and decision rules.
		
		\subsection{Statistical Analysis}\label{sec:stats}
			All analyses were conducted in Python using SciPy \cite{virtanen2020scipy} and statsmodels \cite{spearman1961proof}. We use $\alpha = .05$, report exact $p$ values, and apply no correction for multiple comparisons.
			
			We compared tier distributions across weeks and phases using Pearson's $\chi^2$ test of independence \cite{pearson1900on_the_criterion}, applied to the week-by-tier ($10 \times 6$) and phase-by-tier ($3 \times 6$) contingency tables of prompt counts. We report Cramer's $V = \sqrt{\chi^2 / [N(k-1)]}$ as the effect size, where $N$ is the number of prompts and $k$ is the smaller table dimension \cite{cramer1999mathematical}. To identify the cells contributing most strongly to an association, we report Pearson residuals, $r_{ij} = (O_{ij} - E_{ij}) / \sqrt{E_{ij}}$, where $O_{ij}$ and $E_{ij}$ are the observed and expected counts, respectively. Absolute residuals above approximately 2 indicate cells that depart notably from independence \cite{agresti2013categorical}.
			
			We tested monotonic changes in each tier's weekly share using Spearman's rank correlation $\rho$ between week number and the tier's share in that week \cite{spearman1961proof}. With only ten weekly observations, these tests have low power and describe changes in aggregate weekly shares rather than individual prompts.
			
			We compared the proportion of T0 prompts between deadline and other weeks using a $2 \times 2$ $\chi^2$ test with Yates' continuity correction \cite{yates34contingency}, and report the sample odds ratio (OR) as the effect size \cite{agresti2013categorical}.
			
            The analyses above treat prompts as independent observations, although prompts are nested within students and students contribute different numbers of prompts. We therefore re-estimated the key effects using generalised estimating equations (GEE) \cite{liang86longitudinal}. Each model uses a binary outcome indicating whether a prompt is T0, T4, or T5, with week type or phase as the predictor, an exchangeable working correlation within students, and robust (sandwich) standard errors. We report exponentiated coefficients as odds ratios with 95\% confidence intervals and draw inferential conclusions only for effects that remain supported after accounting for within-student clustering. Clustered checks were limited to the tiers underlying our main claims (T0, T4, T5); other tier shifts are reported descriptively.

			For the longitudinal subset (Section \ref{sec:within-student}), we estimated the proportion of variation in tier attributable to differences between students rather than variation within students using a linear mixed-effects model with a random intercept for student and phase as a fixed effect \cite{laird82}. Tier was treated as numeric for this analysis. The intraclass correlation coefficient was calculated as $\mathrm{ICC} = \sigma^2_b / (\sigma^2_b + \sigma^2_w)$, where $\sigma^2_b$ is the between-student variance and $\sigma^2_w$ is the within-student residual variance \cite{snijders2012multilevel}. For each student, we also calculated the proportion of T0 prompts in deadline and other weeks and compared these paired proportions using the Wilcoxon signed-rank test \cite{wilcoxon45}, excluding students for whom the two proportions were equal.
	
	\section{Results}\label{sec:results}
		Across 962 coded prompts, 84.2\% showed evidence of student engineering contribution (T1-T5). Agency shifted across phases of the SDLC, and these differences remained after accounting for the clustering of prompts within students. T0 was more prevalent during deadline weeks in the unadjusted analysis, but this effect did not remain after accounting for within-student clustering. Most variation in agency occurred within students rather than between students.
		
		Sections \ref{sec:overall-dist} and \ref{sec:delegation:assess:deadlines} analyse all 962 coded prompts. Section \ref{sec:within-student} examines within-student trajectories for the 27 students who submitted an AI Journal in at least 50\% of the eligible weeks, comprising 608 prompts.
		\subsection{Overall Distribution}\label{sec:overall-dist}
			Table \ref{tab:overall-dist} reports tier counts by week. Of the 962 prompts, 152 (15.8\%) were classified as T0, meaning that 84.2\% showed evidence of at least some student engineering contribution (T1--T5). The most frequent tiers were T1 (241; 25.1\%) and T5 (227; 23.6\%), followed by T2 (172; 17.9\%), T4 (119; 12.4\%), and T3 (51; 5.3\%). T3 was the least frequent tier in every week, except Week 9, when it was tied with T5. Its weekly share ranged from 2.7\% in Week 4 to 8.2\% in Week 13, compared with 5.3\% overall. Thus, students relatively rarely specified the system constraints that an artefact had to satisfy when requesting its generation.
			\begin{table}[t] 
				\caption{Weekly distribution across categories T0-T5.} 
				\label{tab:overall-dist} 
				\centering 
				\small 
				\setlength{\tabcolsep}{3.5pt} 
				\begin{tabular}{@{}crrrrrrr@{}} 
					\toprule 
					Week & T0 & T1 & T2 & T3 & T4 & T5 & Total \\ 
					\midrule 
					4 & 28 & 21 & 31 & 5 & 12 & 90 & 187 \\ 
					5 & 19 & 18 & 27 & 8 & 16 & 42 & 130 \\ 
					6 & 28 & 40 & 8 & 7 & 9 & 27 & 119 \\ 
					7 & 19 & 42 & 14 & 9 & 10 & 30 & 124 \\ 
					8 & 8 & 22 & 17 & 3 & 13 & 12 & 75 \\ 
					9 & 7 & 35 & 14 & 3 & 4 & 3 & 66 \\ 
					10 & 6 & 18 & 18 & 3 & 11 & 4 & 60 \\ 
					11 & 10 & 23 & 14 & 4 & 14 & 5 & 70 \\ 
					12 & 13 & 9 & 14 & 3 & 12 & 7 & 58 \\ 
					13 & 14 & 13 & 15 & 6 & 18 & 7 & 73 \\ 
					\midrule 
					\textbf{Total} & \textbf{152} & \textbf{241} & \textbf{172} & \textbf{51} & \textbf{119} & \textbf{227} & \textbf{962} \\ 
					\textbf{\%} & 15.8 & 25.1 & 17.9 & 5.3 & 12.4 & 23.6 & 100 \\ 
					\bottomrule 
				\end{tabular} 
			\end{table}
			\begin{figure*}
				\centering
				\includegraphics[width=.9\textwidth]{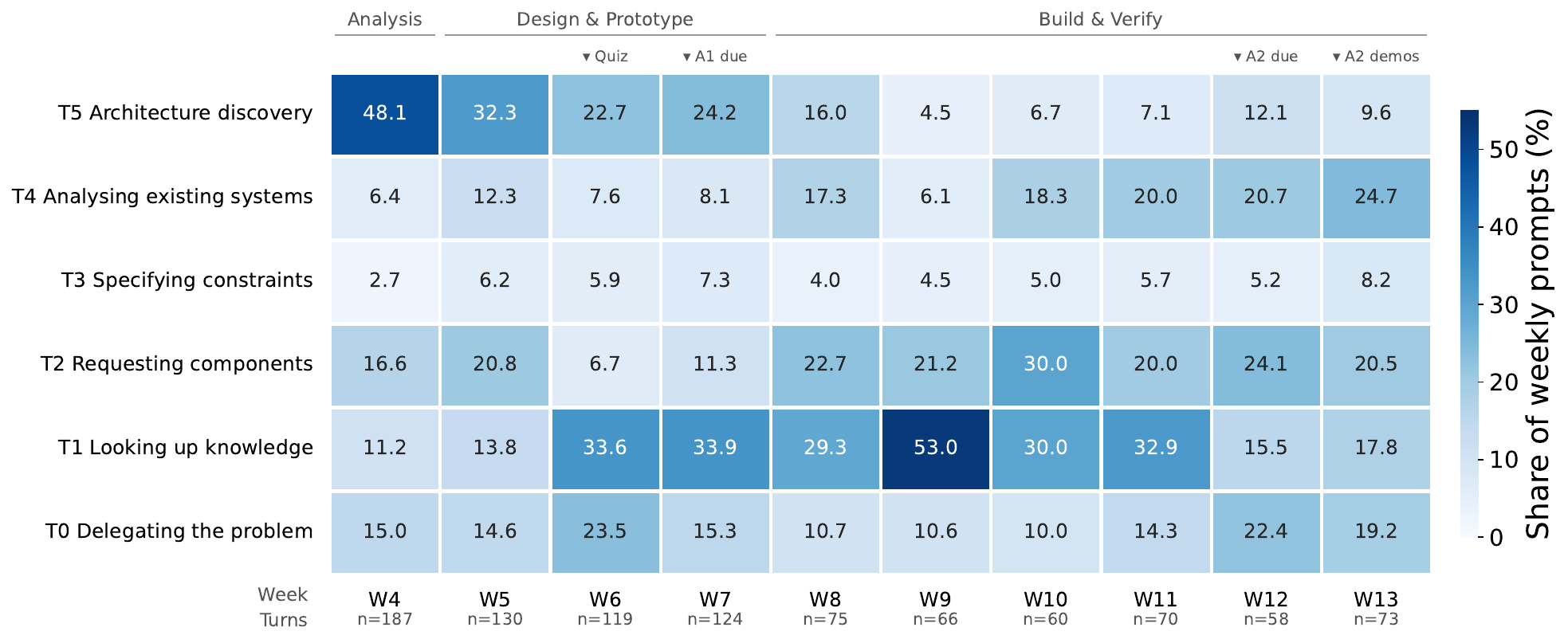}
				\caption{Distribution of HADT tiers by week (962 prompts, weeks 4-13). Each cell is the percentage of that week's prompts at a given tier; columns sum to 100\%, and $n$ is the number of prompts per week. Brackets mark the analysis phases (Table \ref{tab:weekly-schedule}), and $\blacktriangledown$ marks assessment deadlines. Mid-semester break was between weeks 6 and 7.}
				\label{fig:pdc_relative}
			\end{figure*}
			The tier distribution differed across weeks ($\chi^2(45) = 209.8$, $p < .001$, Cramer's $V = 0.21$). Table \ref{tab:weekly-schedule} summarises the course content by week. Based on the development activities in each week, we grouped Weeks 4-13 into three analysis phases: \emph{analysis} (Week 4: requirements engineering and development of the A1 requirements), \emph{design and prototype} (Weeks 5-7: architecture, prototyping, implementation, and the A1 submission), and \emph{build and verify} (Weeks 8-13: verification and validation (V\&V), quality engineering, CI/CD, and deployment while students developed A2). Tier distributions differed more strongly across these phases ($\chi^2(10) = 134.9$, $p < .001$, Cramer's $V = 0.26$). Standardised residuals for the phase-by-tier analysis are shown in brackets:
			\begin{description}
				\item[Analysis: design-space exploration.] T5 accounted for 48.1\% of prompts ($+6.9$) during Week 4, when the course covered requirements engineering and students developed their A1 requirements. Knowledge lookup was relatively uncommon (T1: 11.2\%; $-3.8$).
				\item[Design and prototype: a transition.] T5 remained substantial (26.5\%), while T1 rose to 26.8\%. The largest change occurred in Week 6, when T1 increased from 13.8\% to 33.6\%. At this point, the course moved from prototype design in Figma to implementation using Microsoft Power Apps, SharePoint lists, and AI-assisted coding, while students built their A1 prototypes.
				\item[Build and verify: diagnosis and components.] T4 increased to 17.9\% ($+3.2$) and T2 to 22.9\% ($+2.4$), while T5 fell to 9.5\% ($-5.8$). T1 peaked at 53.0\% in Week 9, when the course covered quality engineering and CI/CD.
			\end{description}
			At the week level, the share of T5 decreased over the semester (Spearman $\rho = -0.76$, $p = .011$, $n = 10$ weeks), while the share of T4 increased ($\rho = 0.78$, $p = .008$, $n = 10$ weeks). Spec-driven development with AI was introduced in Week 12, followed by the highest observed share of T3 in Week 13 (8.2\%). However, Week 13 contained only six prompts, so we do not interpret this pattern further. The number of submitted prompts also declined over the semester, from 187 in Week 4 to 58-75 per week from Week 8 onwards. Consequently, percentage estimates for later weeks are based on fewer observations.
			
			Because prompts are nested within students, we re-estimated the two main phase-related shifts using logistic GEE, accounting for within-student clustering. Both remained significant. Compared with the analysis phase, the odds of a T5 prompt were $0.38$ times as high in the design and prototype phase (95\% CI $0.25-0.56$) and $0.11$ times as high in the build and verify phase (95\% CI $0.05-0.22$; both $p < .001$). Conversely, the odds of a T4 prompt were 2.88 times as high in the build and verify phase (95\% CI $1.41-5.86$, $p = .004$). The increase in the design and prototype phase was smaller and not statistically significant (OR $1.56$, 95\% CI $0.87-2.80$, $p = .14$).
		
		\subsection{Delegation around Assessment Deadlines}\label{sec:delegation:assess:deadlines}
			Four journalled weeks contained assessment deadlines (Table \ref{tab:weekly-schedule}): Week 6 (mid-semester quiz), Week 7 (A1 due), Week 12 (A2 due), and Week 13 (A2 demonstrations and final deliverables). At the prompt level, T0 accounted for 19.8\% of prompts in deadline weeks (74/374), compared with 13.3\% in the other six weeks (78/588; $\chi^2(1) = 6.82$, $p = .009$, OR $= 1.61$). Week 6 had the highest T0 share of the semester (23.5\%). The difference was also pronounced at the end of the semester, with T0 accounting for 20.6\% of prompts in Weeks 12-13 compared with 11.4\% in Weeks 8-11.
			
			However, the deadline effect did not remain statistically significant after accounting for within-student clustering. Because prompts are nested within students, students who submitted more prompts during deadline weeks contributed disproportionately to the prompt-level comparison. In the GEE model, the estimated deadline effect was smaller and not statistically significant (OR $= 1.28$, 95\% CI $0.93-1.75$, $p = .13$). We therefore treat the apparent increase in delegation during deadline weeks as suggestive rather than conclusive. Section \ref{sec:within-student} examines this pattern within students.
					
		\subsection{Within-Student Variation}\label{sec:within-student}
			The longitudinal subset comprised 27 students and 608 prompts. Individual students used a wide range of HADT tiers. The median student used five of the six tiers (range 2-6), and 22 of the 27 students used at least four tiers. Twenty-one students had at least one T0 prompt, but T0 was the most frequent tier for only six.
			
			In a random-intercept model adjusted for phase, 32\% of the variance in tier was attributable to differences between students ($\mathrm{ICC} = 0.32$), while the remaining 68\% occurred within students. Thus, most variation in observed agency occurred within the same students across prompts, although differences between students remained substantial.
			
			The SDLC shift was also visible within individual students. From each student's earliest active phase to the build and verify phase, the share of T5 decreased for 20 of the 27 students, the share of T4 increased for 11, and both changes occurred for 10.
		
			In contrast, the deadline effect was not supported within students. Twelve students had a higher T0 share in deadline weeks than in other weeks, eight had a lower share, and seven showed no change. Although the median T0 share was higher in deadline weeks (16.7\% versus 8.3\%), the paired difference was not statistically significant (Wilcoxon signed-rank $p = .94$). Only two students used T0 exclusively during deadline weeks. A GEE model fitted to the longitudinal subset produced a similar, non-significant estimate (OR $= 1.36$, 95\% CI $0.98$--$1.89$, $p = .066$).
	
	\section{Discussions and Lessons Learnt}\label{sec:discuss}
		The results suggest that first-year students' agency with AI is neither uniformly low nor fixed. Instead, patterns of agency vary with the work students undertake at different stages of the SDLC. Although delegation was more common during deadline weeks in the unadjusted analysis, this pattern was not supported after accounting for repeated prompts within students.
		
		\paragraph{Agency Is Phase-Appropriate, Not a Ladder.} 
        The largest shifts in agency followed the SDLC. Design-space exploration (T5) dominated the early requirements and design phase, whereas diagnosis of existing systems (T4) became more common once students had systems to diagnose. A shift towards lower-numbered tiers later in the semester therefore does not necessarily indicate declining engagement. A T4 prompt during build and verification (``here is my error and what I think causes it'') can demonstrate substantial student agency for that phase, just as a T5 prompt can during requirements and design. HADT tiers are therefore better interpreted as a profile in relation to the phase of engineering work than as a score to maximise. In practice, educators could identify the tier profiles expected at different stages of a project and use HADT to examine where the observed profile of a cohort differs from those expectations.
		
		\paragraph{Mid-Semester Knowledge Lookup.} 
        Factual and procedural lookup (T1) rose sharply in Week 6 and peaked at 53\% of prompts in Week 9. The rise began as the course moved from prototyping in Figma to implementation in a low-code/no-code environment using Power Apps and SharePoint lists, tools with which most students had little prior experience. Using AI as a reference for platform syntax, formulas, and features is a legitimate form of assistance, and T1 is not equivalent to delegation. However, repeated lookup can become problematic if it substitutes for developing the underlying knowledge needed to evaluate and reuse the answers. For a course such as this, a checkpoint on platform fundamentals may therefore be more appropriate than discouraging students from using AI for lookup.
		
		\paragraph{Agency Varies Within Students; Deadlines Are a Weak Driver.} 
        T0 accounted for between 10\% and 24\% of prompts each week, so most prompts in every week showed evidence of student engineering contribution. The observed level of agency was also not simply a stable characteristic of individual students. In the phase-adjusted model, 32\% of the variance in tier lay between students and 68\% within students, while the median student used five of the six tiers over the semester. These results suggest that the same student can adopt substantially different levels and forms of agency across different prompts and stages of work.
		
		Deadline pressure, however, provided little evidence of explaining this variation. The prompt-level increase in T0 during deadline weeks became smaller and was no longer statistically significant after accounting for within-student clustering. The within-student comparison was similarly inconclusive: 12 students had a higher T0 share during deadline weeks, eight had a lower share, and seven showed no change. In this dataset, SDLC phase is therefore more clearly associated with patterns of agency than the occurrence of a deadline. Assessment design may nevertheless matter through the kinds of engineering activity that particular stages and deliverables require.
		
		\paragraph{Constraints Are Rarely Specified.} 
        T3, in which students specify the system constraints that an artefact must satisfy, was the least frequent tier in every week (5.3\% overall). Students more often requested a component without its wider system constraints (T2) or framed a design space for exploration (T5) than explicitly translated project requirements into constraints on generated artefacts.
		
        This pattern did not arise in the absence of instruction. Students were taught techniques for constraining AI output through examples, output-format specifications, and explicit success criteria at several points in the course. One possible explanation is therefore that T3 depends not only on prompting technique but also on sufficient domain and project knowledge to formulate useful constraints. To specify a meaningful constraint, students may need to know what properties a suitable artefact should have, including the relevant data model, platform limitations, and acceptance criteria. First-year students were developing this knowledge during the semester, so their ability to articulate constraints may have lagged behind their exposure to prompting techniques. This interpretation is consistent with evidence that beginners can struggle to express intent in forms that code LLMs interpret as intended \cite{nguyen2024beginning}. On this interpretation, teaching prompting techniques alone may be insufficient to increase T3 behaviour. However, the prompt data cannot distinguish students who lacked the knowledge needed to specify constraints from those who knew the relevant constraints but chose not to include them.

		\paragraph{Using HADT in Teaching.} 
        Making HADT visible to students could help them recognise the kinds of engineering contribution evident in their prompts and consider whether those contributions are appropriate for their current phase of work. There is, however, a tension between using HADT pedagogically and using AI Journals as records of authentic practice. If particular tiers are rewarded, students may optimise their prompts to satisfy the taxonomy rather than use AI in the way they otherwise would, an instance of the measurement problem associated with Goodhart's law \cite{strathern1997improving}. We therefore recommend using HADT primarily for formative feedback, cohort-level analysis, and course design rather than assigning marks for reaching particular tiers. Keeping the AI Journal itself unmarked may help preserve its role as a record of students' actual AI use.
		
		\paragraph{Transferability.} 
            The three HADT dimensions are not inherently specific to software engineering. Contextual bandwidth, level of engineering abstraction, and problem ownership can arise wherever students use GenAI to produce or refine disciplinary artefacts. The specific manifestations of these dimensions, however, are likely to be discipline-dependent. HADT may therefore provide a basis for adaptation to other computing and engineering contexts, but its tier definitions and decision rules should be validated before being transferred to substantially different tasks or disciplines.
	
	\section{Limitations and Threats}\label{sec:limits}
		HADT measures the engineering contribution made visible in each prompt, not agency across an entire conversation. Context carried forward from earlier prompts or retained in model memory is therefore not credited. This is a deliberate construct choice. Coders used only the prompt text: attachments were not recorded in the AI Journal, and other journal fields were not consulted. Constraints or contextual information supplied outside the prompt are therefore invisible to HADT. The tiers are ordinal but do not form a quality scale (Section \ref{sec:discuss}). Analyses that treat tier as numeric, particularly the ICC analysis, should therefore be interpreted as describing the distribution of tiers rather than better or worse prompting. 
		
        The AI Journal is self-reported, i.e. students chose which AI interactions to record or they could instead declare that they had not used AI. Although the journals were unmarked and students were explicitly told that their disclosures would not reduce their marks, selective reporting remains possible. Delegated interactions may be particularly susceptible to under-reporting, so the observed T0 proportion should not be interpreted as the true prevalence of delegation. Journal participation also declined over the semester, from 187 prompts in Week 4 to 58-75 per week from Week 8 onwards. Later weeks therefore contain fewer observations and may represent a different subset of students from earlier weeks, potentially biasing comparisons across the semester. The clustered models and longitudinal subset reduce some consequences of unequal contributions across students but cannot eliminate selection arising from which students chose to submit journals or which interactions they recorded.

		The SDLC phases are also confounded with the course activities and instruction delivered during each period (Table \ref{tab:weekly-schedule}). Lecture topics, explicit instruction in AI use (e.g., using LLMs for requirements engineering in Week 4), project activities, and assessment demands all changed alongside the lifecycle phase. We therefore cannot isolate whether the observed differences in agency are associated with the engineering phase itself, the accompanying instruction, the assessment context, or some combination of these factors.

        The prompt-level $\chi^2$ analyses treat prompts as independent even though multiple prompts are nested within students. We therefore report GEE estimates that account for within-student clustering alongside these analyses and base our main inferential conclusions on effects that remain supported after clustering is taken into account. The apparent increase in T0 during deadline weeks did not meet this criterion and is therefore treated as suggestive rather than conclusive. Several statistical tests were also conducted on the same dataset without correction for multiple comparisons, increasing the risk of false-positive findings. Conversely, the week-level trend tests contain only ten observations, and T3 counts are small throughout the semester, limiting power to detect small effects and making week-specific T3 patterns particularly unstable.
		
        Finally, the data come from a single first-year course in one semester at one university. Students had no assumed prior programming experience, implemented their projects in a low-code/no-code environment (Power Apps with SharePoint lists) rather than through conventional programming, and were explicitly taught how to use GenAI. Agency profiles may differ in programming-intensive courses, among more experienced students, in other disciplinary contexts, or where AI use is permitted but not explicitly taught. The longitudinal analysis further includes only the 27 students who submitted an AI Journal in at least half of the eligible weeks, compared with 105 students who contributed at least one analysed journal. This subset may not be representative of all students who used AI or of the cohort as a whole. The longitudinal findings should therefore be interpreted as evidence about variation within this more consistently observed subset rather than as estimates of within-student behaviour for the entire cohort.

	\section{Conclusions and Future Research Directions}\label{sec:conc}
		When AI produces the artefact, the cognitive complexity of the task no longer tells educators what the student contributed. We therefore asked a different question: ``who is doing the thinking?'' We proposed HADT, a six-tier taxonomy that classifies the engineering contribution evident in each prompt along three dimensions: contextual bandwidth, engineering abstraction, and student agency. Applied to 962 prompts from a first-year, AI-permissive software engineering course, HADT showed that most prompts contained evidence of student engineering contribution. Patterns of agency shifted across the development lifecycle, from design-space exploration during requirements to diagnosis during build and verification, and varied more within students than between them. Delegation did not increase reliably during deadline weeks after accounting for within-student clustering, while explicit specification of system constraints remained rare despite being taught.
		
		For educators, the central message is that appropriate agency depends on the engineering work students are undertaking. This reinforces the need to redesign AI-permissive assignments carefully around the contributions students are expected to make, rather than simply around the artefacts they produce. HADT provides a lightweight way to make those contributions visible through a tier profile for each week or project phase, interpreted against the engineering work that phase requires rather than as a score to maximise.
		
		Future work should refine and validate HADT across courses, institutions, experience levels, and programming contexts. A second direction is to investigate whether coding can be scaled through LLM-assisted classification validated against human labels. Reliable automated coding could allow HADT profiles to support timely formative feedback during a semester, for example through an AI Journal dashboard.

\begin{acks}
	This project was supported by an Education-Focused Academics SoTL Grant from the University of Sydney. 
\end{acks}

\section*{Ethics and Privacy Statement}
    This study was approved by the Human Research Ethics Committee of the University of Sydney (Approval No. HREC 2025/HE001132-substudy 022 v1a). Journal data were de-identified prior to analysis and reported only in aggregated form. Participation formed part of normal course activities; however, all research analyses were conducted on anonymised data in accordance with institutional ethics requirements.

\bibliographystyle{ACM-Reference-Format}
\bibliography{references}

%
%
%
%
%
%
%
%

\end{document}